# Current induced generation and synchronous motion of highly packed coupled chiral domain walls

Rafael P. del Real[1], Victor Raposo[2], Eduardo Martinez[2] and Masamitsu Hayashi[3,4*]

[1]*Instituto de Ciencia de Materiales de Madrid, CSIC, 28049, Spain*

[2]*University of Salamanca, Plaza de los Caidos s/n, E-37008 Salamanca, Spain*

[3]*Department of Physics, The University of Tokyo, Bunkyo, Tokyo 113-0033, Japan*

[4]*National Institute for Materials Science, Tsukuba 305-0047, Japan*

ABSTRACT: Chiral domain walls of Neel type emerge in heterostructures that include heavy metal (HM) and ferromagnetic metal (FM) layers owing to the Dzyaloshinskii-Moriya (DM) interaction at the HM/FM interface. In developing storage class memories based on the current induced motion of chiral domain walls, it remains to be seen how dense such domain walls can be packed together. Here we show that a universal short range repulsion that scales with the strength of the DM interaction exists among chiral domain walls. The distance between the two walls can be reduced with application of out of plane field, allowing formation of coupled domain walls. Surprisingly, the current driven velocity of such coupled walls is independent of the out of plane field, enabling manipulation of significantly compressed coupled domain walls using current pulses. Moreover, we find that a single current pulse with optimum amplitude can create a large number of closely spaced domain walls. These features allow current induced generation and synchronous motion of highly packed chiral domain walls, a key feature essential for developing domain wall based storage devices.

Spin orbit effects[1-3] in magnetic heterostructures have altered the conventional approach to controlling magnetization electrically. The spin Hall effect[4-6] in non-magnetic heavy metals (HM) can generate large enough spin current to cause switching of the magnetization of a neighboring magnetic layer[2]. Recent studies have shown that such diffusive spin current from the spin Hall effect can also drive domain walls[7-13] of neighboring magnetic layer if the walls adopt a chiral Neel configuration. The Dzyaloshinskii-Moriya (DM) interaction[14-16] at the interface of HM layer and a ferromagnetic metal (FM) layer enables formation of such chiral Neel walls[17-24]. In many cases, the spin Hall angle of the HM layer and the strength of the DM interaction define the efficiency of current induced motion of chiral domain walls.

In developing domain wall based devices[25-27], one of the main challenges that needs to be assessed is its scalability. In particular, it remains to be seen how dense domain walls can be packed within a quasi-one dimensional wire. Recently it has been reported that a topological repulsive force exists among vortex (and in some cases transverse) domain walls in in-plane magnetized systems[28]. Short range repulsive interaction between domain walls may be advantageous for placing domain walls close to each other, allowing dense packing of domain walls[29, 30]. However, the repulsive interaction among the vortex (and transverse) domain walls depends on the successive alignment of magnetic charges of neighboring domain walls. Thus the repulsive force may be absent for particular alignments.

Here we show that a universal repulsive interaction exists among chiral Neel walls. The repulsive force increases with increasing strength of DM interaction. Due to the short range repulsive interaction, two neighboring walls can form a coupled state in the presence of an out of plane field. Whereas current driven velocity of an isolated wall scales linearly with the out of plane field, the velocity of the coupled walls shows little dependence on the field. This allows one to drive coupled domain walls that are significantly compressed using current pulses. In addition, we find that current pulses can generate a large number of closely spaced domain walls when appropriate pulses are used. These features enable generation and synchronous motion of highly packed domain walls using current pulses.

Films are grown on Si substrates using magnetron sputtering. The film structure is Sub.|1.5 Ta|$d$ W|1 CoFeB|2 MgO|1 Ta (units in nanometer). The thickness of the W insertion layer ($d$) is varied to control the DM exchange interaction[12], as demonstrated below. Optical lithography and Ar ion milling are used to form a wire, typically ~5 μm wide and ~40 μm long. Figure 1(a) shows a schematic illustration of the film stack and a Kerr microscopy image of a representative device. Magnetic properties of the films are shown in Fig. 1(b-e). The W layer thickness dependence of the saturated magnetic moment per unit volume ($M/V$) and the effective magnetic anisotropy energy ($K_{EFF}$), obtained from magnetization hysteresis loops, are shown in Figs. 1(b) and 1(c). Whereas $M/V$ is almost constant against $d$, $K_{EFF}$ increases with increasing $d$ and saturates when $d$~0.5 nm.

The inset of Fig. 1(d) shows the voltage pulse amplitude dependence of the current driven domain wall velocity ($v$). The velocity saturates for large pulse amplitudes. The saturation velocity ($v_D$) is plotted as a function of $d$ in Fig. 1(d). According to the one-dimensional (1D) model[7, 31] of domain walls, $v_D$ is proportional to the DM exchange constant $D$, i.e. $v_D = \gamma D/M_S$, where $\gamma$ is the gyrotropic ratio. $M_S$ is the saturation magnetization of the magnetic layer and here we assume it is equivalent to $M/V$. The DM exchange constant is calculated using the above relation and its $d$-dependence is shown in Fig. 1(e). $D$ increases with increasing $d$ till $d$~0.6 nm, above which it tends to saturate.

We next show current induced nucleation of domain walls. In the wires studied, we find that a large number of domain walls can be formed under certain conditions. Figure 2(a) shows the pulse length dependence of the number of domain walls nucleated upon application of a single voltage pulse. The inset of Fig. 2(a) shows representative magnetic states of the wire after application of the voltage pulse. For larger pulse amplitudes (~±35 V, ~±40 V), the number of domain walls shows little dependence on the pulse length: the number lies in between ~5 to ~10. In contrast, a sharp increase in the number of domain walls with the pulse length is found when the amplitude is set to ~±32 V. Below ~±32 V, we find little evidence of current induced nucleation of domain walls for this device.

The W insertion layer thickness dependence of the average number of domain walls nucleated when the pulse length is fixed to 100 ns is shown in Fig. 2(b). The pulse amplitude is set such that the number of domain walls nucleated is at maximum. Interestingly, we find that the domain wall nucleation significantly depends on $d$. When $d<\sim0.5$ nm, the number of domain walls nucleated is fixed to a value less than ~10. The inset of Fig. 2(b) enclosed by the blue rectangle shows a representative image of the magnetic state. For wires with $d<\sim0.5$ nm, we find that the wall normal of the nucleated domain walls is not always directed along the wire's long axis. The number of domain walls abruptly increases when $d$ exceeds ~0.5 nm. Although there seems to be a correlation between the number of domain walls nucleated and the strength of DM interaction, we find that a non-zero $D$ is not a sufficient condition in order to observe large number of domain walls nucleated by voltage pulses. Further investigation is required to clarify the underlying mechanism of this effect.

Nucleation of multiple walls allows us to study the interaction among them using out of plane magnetic field ($H_Z$). We start from the multi-domain state created by the voltage pulse application. To study the interaction, Kerr images are taken at near zero field before and after $H_Z$ is changed to a certain value. The number of domain walls existing in the wire after application of $H_Z$ is plotted in Fig 3(a). The corresponding magnetic states of the wire, obtained by Kerr imaging at near zero field, are shown in Fig. 3(b). Here the field is applied such that it compresses the width of the bright domains (magnetization pointing along +z) which are nucleated by the voltage pulse. At small fields ($|H_Z|<\sim10$ Oe), the width of bright domains decreases with increasing $|H_Z|$. The mean $|H_Z|$ at which the domain walls move is defined as the propagation field ($H_P$). When $|H_Z|$ exceeds $H_P$, the width of each domain remains constant as the Kerr images here are captured at near zero field. Note that Kerr images taken at large $|H_Z|$ show less contrast of the compressed domains, suggesting that the domain width reduces with increasing $|H_Z|$. Pairs of domain walls start to annihilate one another when $|H_Z|>\sim20$ Oe. The last two walls collapse at $|H_Z|\sim37$ Oe. The mean and maximum values of the annihilation field $H_{AN}$ are defined as the $|H_Z|$ at which the number of domain walls is reduced to half of the initial state and to zero, respectively.

The W layer thickness dependence of $H_P$, the mean and the maximum $H_{AN}$ are shown in Fig. 3(c). Except for the abrupt increase of $H_P$ and $H_{AN}$ when $d$ approaches ~1 nm, whose origin is not clear, the thickness dependence of $H_P$ and $H_{AN}$ is different: whereas $H_P$ tends to decrease with increasing $d$, $H_{AN}$ scales with $d$. The difference indicates that parameters responsible for defining $H_P$ and $H_{AN}$ are different. As studied extensively[32], part of the change in $H_P$ with $d$ is related to the variation of the domain wall width, which scales with $1/\sqrt{K_{EFF}}$. In contrast, the thickness dependence of $H_{AN}$ may be related to the variation of $D$: both $H_{AN}$ and $D$ scale with $d$ in a similar way (except when $d$ is close to ~1 nm). We find nearly a ~60% increase in $H_{AN}$ when comparing wires with $D$ of ~0 ($d\sim0.1$ nm) and ~0.25 erg/cm$^2$ ($d\sim0.6$ nm). This is in accordance with micromagnetic simulations, which predict a similar increase in $H_{AN}$ with $D$ (see Supporting Information S1). These results demonstrate that there is a strong repulsive force among the domain walls with large $D$.

We now show that the repulsive force between two neighboring domain walls and application of an out of plane field result in formation of a coupled wall. The open symbols of Fig. 4(a) display the current driven velocity of an isolated (single) domain wall as a function of $H_Z$ applied during the voltage pulse application. The velocity ($v$) shows a significant dependence on $H_Z$. Close to zero $H_Z$, we can fit the results with a linear function, which are plotted by the solid lines in Fig. 4(a). According to the 1D model of a domain wall (see Supporting Information S2), the slope of $v$ vs. $H_Z$ is inversely proportional to the current induced effective field ($H_{SH}$) that arises from the spin orbit torque at the HM/FM interface. The variation of $v$ with $H_Z$ is inevitable for domain walls driven by the spin Hall effect of the HM layer and is not preferable for technological applications.

Figure 4(b) shows successive Kerr images of the current driven domain wall motion under the application of $H_Z$ (~-6.5 Oe). As the velocity for ↑↓ and ↓↑ walls are different due to the out-of-plane field, the two walls initially approach each other (panels i to iv). Interestingly, once the two walls merge to form a coupled state, the walls move together along the current flow (panels v to viii). This indicates that the velocities of the ↑↓ and ↓↑ walls are the same once they form the coupled state. The velocities of the coupled domain walls are plotted in Fig. 4(a) using the solid symbols. We find that velocity of the coupled state is approximately the same with that of a single wall with $H_Z$=0 and that it shows little dependence on $H_Z$. Note that the wall no longer forms a coupled state when $H_Z$ is positive for the configuration shown in Fig. 4(a) and 4(b): positive $H_Z$ here promotes separation of the two walls and the walls move with different velocity due to the different net force each wall experiences from the current and field (see Supporting Information S1 and S2 for the details).

Micromagnetic simulations are performed to account for the results shown in Fig. 4 (see Supporting Information S1). When the walls form a coupled state, $H_Z$ is compensated by the repulsive dipolar field from the neighboring domain wall. That is, $H_Z$ sets the distance between the two domain walls such that the two walls experience net zero field. Thus the current driven velocity of the coupled walls under non-zero negative $H_Z$ remains the same with that of an isolated wall at zero field. The distance between the two walls that form the coupled state decreases with increasing $|H_Z|$: the distance decreases by nearly one decade when $|H_Z|$~20 Oe is applied compared to that at $|H_Z|$=0 Oe. The simulations indicate that it is possible to move coupled domain walls, placed as close as ~50 nm, with current pulses owing to the dipolar field between the magnetic moments of chiral Neel walls. We note that the repulsive force between the chiral Neel walls increases with increasing $D$, allowing further reduction of the separation distance (see Supporting Information S1).

To this end, we demonstrate, as shown in Figs. 5(a-c), current induced synchronous motion of highly packed coupled domain walls (number of domain walls is ~18-24). Here we use a wire with $d$~0.6 nm so that a large number of domain walls can be nucleated with the application of a long (~100 ns) voltage pulse. At near zero field, all walls move synchronously along the current flow direction (Fig. 5(b)). When a negative out of plane field that compresses the bright domains is applied (Figs. 5(a)) neighboring walls will first move in opposite direction (see the open symbols in Fig. 4(a)). However, owing to the repulsive interaction and the formation of the coupled states we find that the domain walls do not annihilate one another and move synchronously, even under the application of $H_Z$. When a positive $H_Z$ is applied, the bright domains are initially expanded and the dark domains become compressed. The coupled walls at $H_Z$~10 Oe also move synchronously with voltage pulses. During this process, however, we find that one coupled state has been annihilated (compare the top and the next image of Fig. 5(c)).

The annihilation of the coupled domain walls during their motion needs to be addressed in particular for device applications. The results shown in Fig. 3 suggest that the distribution of the annihilation field $H_{AN}$ among the ~20 domain walls is quite large. Such distribution may arise either from the spatial variation of pinning or the DM interaction. Recent reports on similar systems have shown that the DM exchange strength may vary locally in a significant way[24]. Such large variation of $D$ along the wire can explain the annihilation[33] of coupled domain walls while they are moving: once the walls enter a region in which $D$ is locally small, the repulsive interaction will reduce and thus may allow easier annihilation. Controlling the uniformity of $D$ may become essential for moving a large number of domain walls with current.

The coupled chiral Neel walls are similar to Skyrmions in the context of magnetic texture[34]: the projection of the magnetization direction along the wire is similar. However, whereas an out of plane magnetic field can control the separation distance between neighboring chiral domain walls allowing formation of highly packed states, the field will only modify the size (diameter) of skyrmions and will not generate densely packed trains of skyrmions. As the repulsive force between the chiral Neel walls increases with increasing DM interaction, the separation distance between the walls can be further reduced. This will allow highly packed coupled domain walls that can be synchronously driven along the wire by current pulses. These results provide new perspectives on utilizing chiral magnetic textures for developing information storage and computation devices.

ACKNOWLEDGMENT

This work was partly supported by JSPS Grant-in-Aid for Specially Promoted Research (15H05702), Grant-in-Aid for Scientific Research (16H03853), MEXT R & D Next-Generation Information Technology. R.P.d.R thanks Subprograma de Movilidad from the Spanish Ministry of Education, Culture and Sports (PRX14/00311). The work by E. M. and V. R. was supported by project WALL, FP7-PEOPLE-2013-ITN 608031 from European Commission, project MAT2014-52477-C5-4-P from Spanish government, and project SA282U14 from Junta de Castilla y Leon.

MATERIALS AND METHODS

A. Kerr images

All Kerr images are subtracted images. The reference is a saturated state with the wire's magnetization pointing along –z. Bright and dark contrast correspond to a magnetic state with magnetization directed along +z and –z, respectively.

B. Velocity measurements

Current driven velocity of domain walls is estimated by dividing the distance the wall traveled with the length of the applied voltage pulse. The distance the wall traveled is calculated using the Kerr images taken before and after the voltage pulse application. For the latter, the Kerr image is captured ~10 ms after the pulse application. The ~10 ms delay provides sufficient time for the wall to relax.

C. Annihilation field measurements

The annihilation field is measured as the following. First, multiple domain walls are created by applying a long (~100 ns) voltage pulse to the wire. A Kerr image is taken to record the initial state. An out of plane field is applied using an electromagnet (rise time ~100 ms). The field is maintained to a constant value for ~200 ms. A Kerr image is taken while the field is on to record the magnetic state under the field. The field is then reduced to near zero and a Kerr image is recorded after the electromagnet settles. We compare the Kerr images at near zero field to study the number of domain walls.

FIGURES

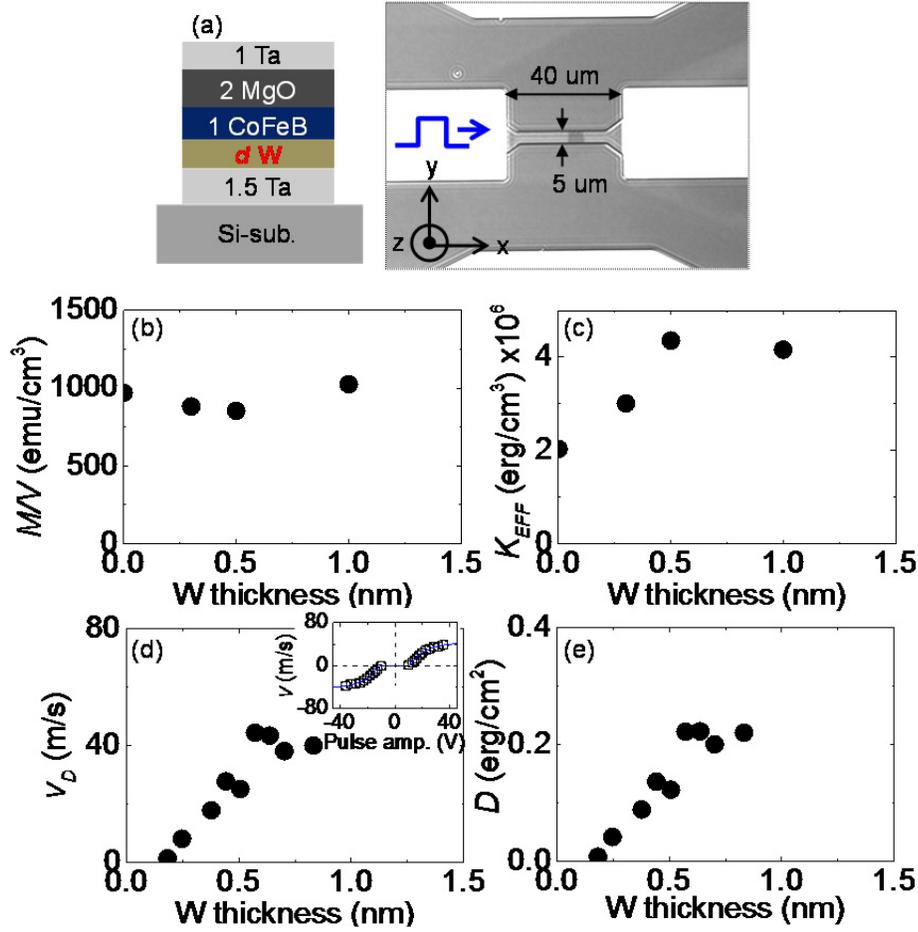

**Fig. 1.** Magnetic properties of the heterostructures studied. (a) Schematic illustration of the film structure and an optical microscopy image of a representative wire used in the experiments. (b-e) Saturated magnetic moment per unit volume $M/V$ (b), the effective magnetic anisotropy energy $K_{EFF}$ (c), the saturation domain wall velocity $v_D$ (d) and the DM exchange constant $D$ (e) plotted as a function of the W insertion layer thickness $d$. The inset of (d) shows the domain wall velocity $v$ as a function of the voltage pulse amplitude for a wire with $d$~0.6 nm.

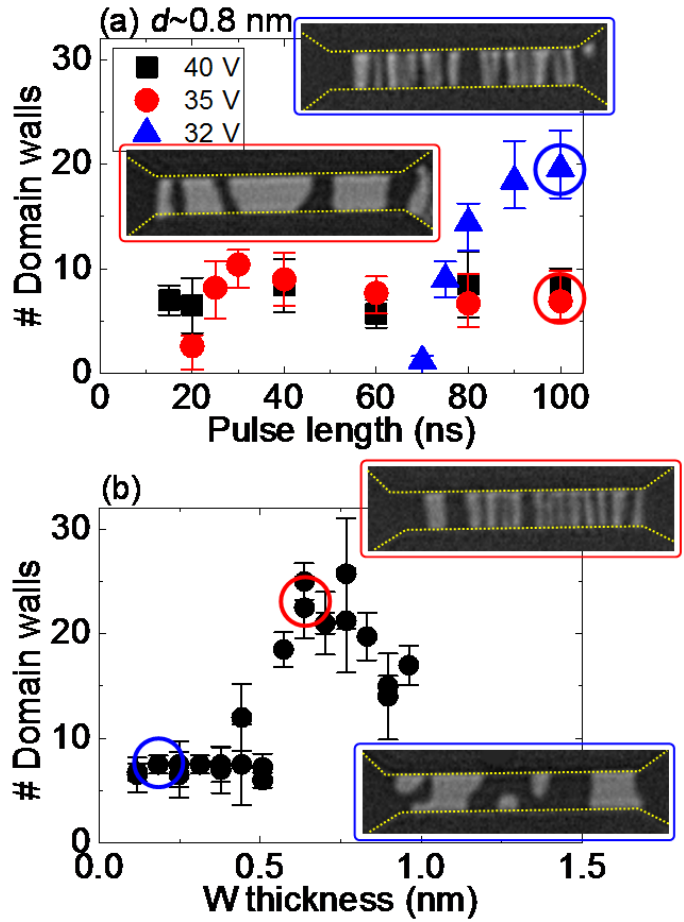

**Fig. 2.** Current induced generation of highly packed domain walls. (a) Voltage pulse length dependence of the average number of domain walls created in a wire with $d\sim0.8$ nm. Symbols note results from different pulse amplitudes. The error bars show standard deviation of the number of domain walls created in 5 successive measurements. The Kerr images show representative magnetic states after a voltage pulse is applied corresponding to the condition marked by the open circles. (b) W insertion layer thickness dependence of the average number of domain walls created by voltage pulses. The pulse amplitude is set to a value that results in generating the largest number of domain walls (i.e. ±32 V or ±35 V). The pulse length is fixed to 100 ns. The error bars show standard deviation of the number of domain walls created in 4

independent measurements. Kerr images: representative magnetic state after a voltage pulse application of a film structure with W insertion layer thickness marked by the open circles.

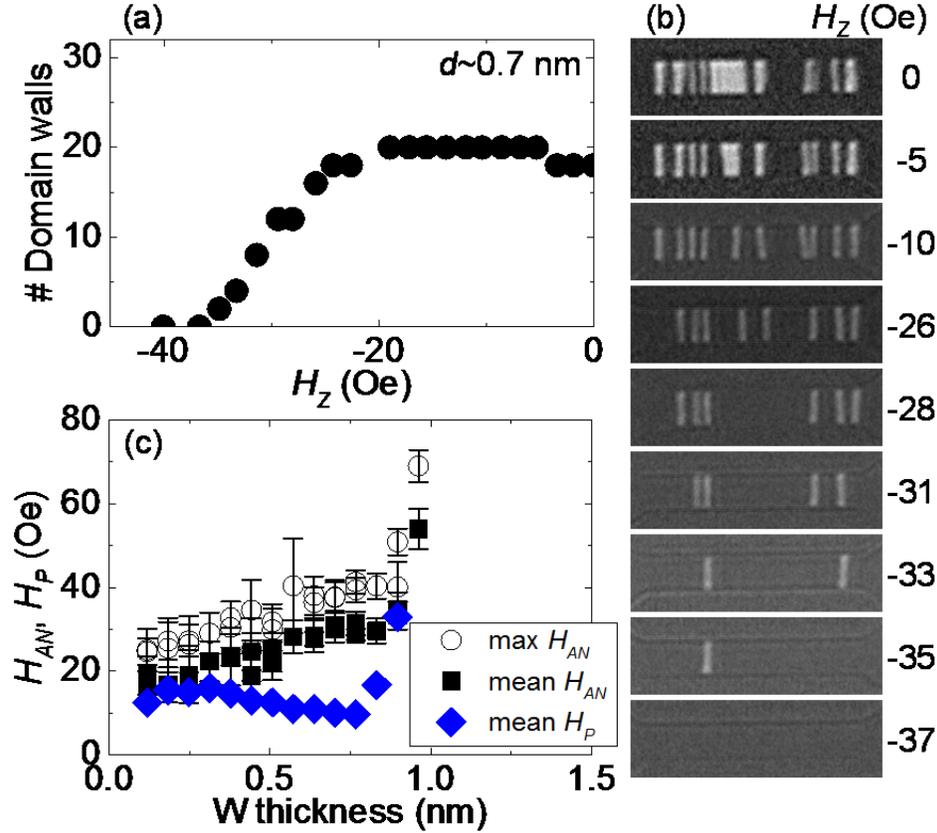

**Fig. 3.** Repulsive interaction and formation of coupled chiral domain walls. (a) Variation of the number of domain walls as a function of out of plane magnetic field ($H_Z$). (b) Snapshots of the magnetic state after application of $H_Z$: the value of $H_Z$ is indicated in the right of each image. The images are captured at near zero field. The sample used in (a) and (b) has W thickness of ~0.7 nm. (c) W thickness dependence of the average $H_Z$ needed to annihilate all (circles) and half of (squares) domain walls. The diamonds show the mean propagation field of domain walls. The error bars show standard deviation of annihilation ($H_{AN}$) and propagation ($H_P$) fields from 4 independent measurements.

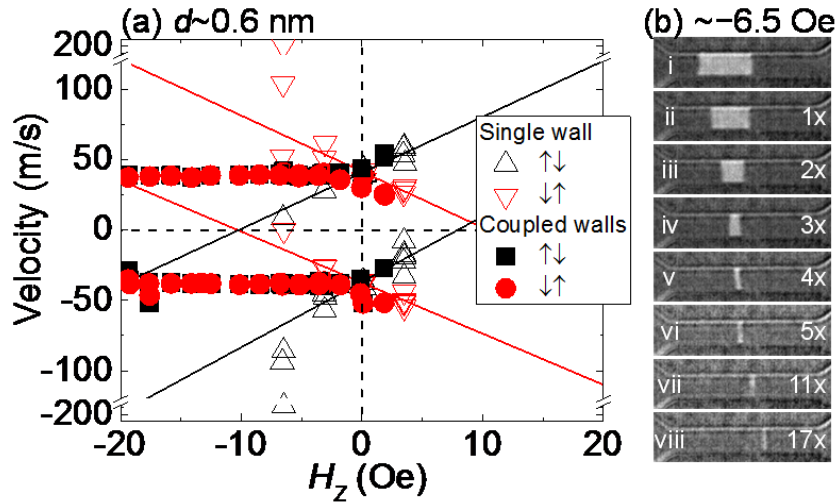

**Fig. 4.** Current driven motion of coupled chiral walls. (a) Domain wall velocity vs. out of plane field ($H_Z$) for isolated (open symbols) and coupled (solid symbols) walls. Squares and up triangles (circles and down triangles) represent the velocity for ↑↓ (↓↑) walls. The line shows linear fit to the velocity of the isolated walls around zero field. (b) Snapshots of the magnetic state of the wire after application of voltage pulses. The number of voltage pulses applied after the initial state is indicated in the right of each image. An out of plane field of ~-6.5 Oe is applied during the pulse application. (a,b) The pulse amplitude and length are ~±32 V and ~10 ns, respectively. The sample used in (a) and (b) has W thickness of ~0.6 nm.

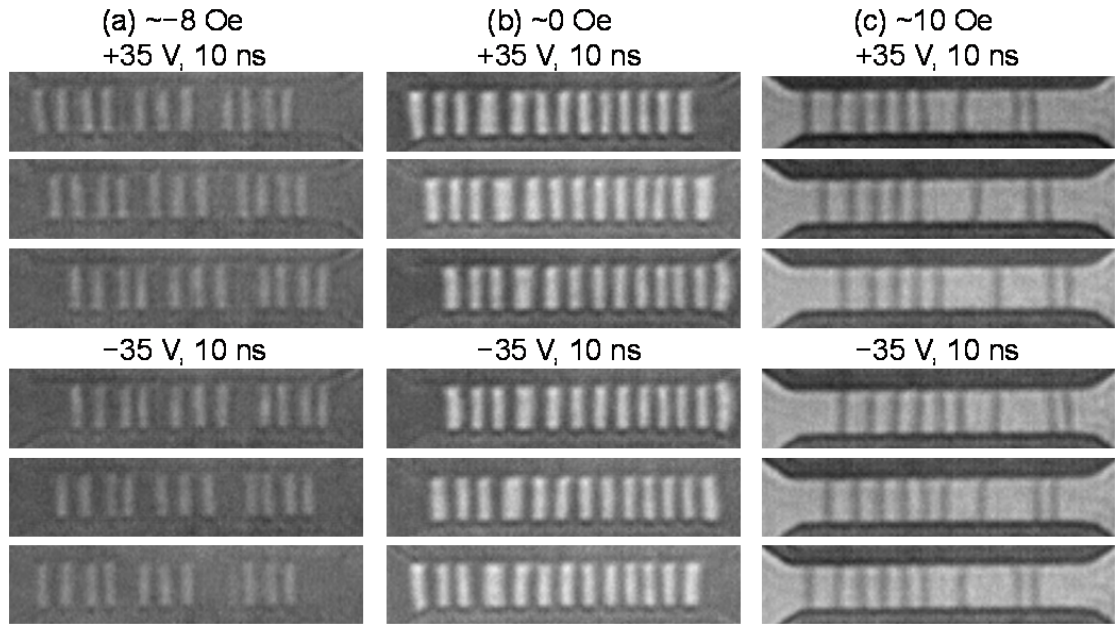

**Fig. 5.** Current controlled motion of highly packed chiral domain walls. (a-c) Snapshots of the magnetic state of the wire after application of voltage pulses. Between each image, five voltage pulses are applied. The pulse amplitude and length are described in the legend. The out of plane field ($H_Z$) is ~-8 Oe (a), ~0 Oe (b) and ~10 Oe (c). The W thickness of the sample used here is ~0.6 nm.

**Supporting Information for**

**Current induced generation and synchronous motion of highly packed coupled chiral domain walls**

Rafael P. del Real[1], Victor Raposo[2], Eduardo Martinez[2] and Masamitsu Hayashi[3,4*]

[1]*Instituto de Ciencia de Materiales de Madrid, CSIC, 28049,Spain*

[2]*University of Salamanca, Plaza de los Caidos s/n, E-37008 Salamanca, Spain*

[3]*Department of Physics, The University of Tokyo, Bunkyo, Tokyo 113-0033, Japan*

[4]*National Institute for Materials Science, Tsukuba 305-0047, Japan*

*Email: hayashi@phys.s.u-tokyo.ac.jp

**Supporting Information**

**S1. Full micromagnetic simulations**

**S2. 1D model for two domain walls**

## S1. Full micromagnetic simulations

### (a) Model description

Micromagnetic (μM) simulations are performed by solving the Landau Lifshitz Gilbert equation augmented with the spin-orbit torque ($\vec{\tau}_{SOT}$)[1-5]:

$$\frac{\partial \vec{m}}{\partial t} = -\gamma \vec{m} \times (\vec{H}_{\text{eff}} + \vec{H}_{\text{th}}) + \alpha \vec{m} \times \frac{\partial \vec{m}}{\partial t} + \vec{\tau}_{SOT} \tag{S1}$$

where the effective field $\vec{H}_{\text{eff}}$ includes Zeeman, exchange, magnetostatic, magnetocrystalline anisotropy (i.e. uniaxial perpendicular magnetic anisotropy) and Dzyaloshinskii-Moriya interactions. $\vec{H}_{\text{th}}$ is the thermal field[6]. The spin-orbit torque ($\vec{\tau}_{SOT}$) includes both damping-like ($\vec{\tau}_{DL}$) and field like ($\vec{\tau}_{FL}$) contributions:

$$\vec{\tau}_{SOT} = \vec{\tau}_{DL} + \vec{\tau}_{FL} = \gamma H_{DL} \vec{m} \times (\vec{\sigma} \times \vec{m}) + \gamma H_{FL} (\vec{\sigma} \times \vec{m}) \tag{S2}$$

where $H_{DL} = \frac{\hbar \theta_{SH} J(t)}{2 e M_S t_{FM}}$ parameterizes the damping-like spin-orbit torque (DL-SOT) in terms of the spin Hall angle ($\theta_{SH}$)[7, 8] and the injected current density ($J(t)$). $t_{FM}$ is the thickness of the ferromagnetic layer. $H_{FL} = k H_{DL}$ parameterizes the field-like spin-orbit torque (FL-SOT), which is quantified by the field-like parameter $k$ representing the ratio between the field-like and the damping-like spin-orbit torques. $\vec{\sigma} = \vec{u}_y$ is the polarization of the spin current (see Refs. 5 and 9 for numerical details) entering the ferromagnetic layer. The definition of $\vec{u}_x$, $\vec{u}_y$ and $\vec{u}_z$ are described in Fig. 1(a) of the main text. Typical material parameters are obtained from experimental measurements[10]: saturation magnetization $M_S = 1100$ emu/cm³, exchange constant $A = 1.5 \times 10^{-6}$ erg/cm, uniaxial anisotropy constant $K_u \sim 1.08 \times 10^6$ erg/cm³, spin Hall angle $\theta_{SH} = -0.21$, and Gilbert damping parameter $\alpha = 0.05$. The DM exchange constant and the FL-SOT are $D = 0.24$ erg/cm$^2$ and $k = 0.5$ respectively, unless noted otherwise.

In order to take into account the effects of disorder due to imperfections and defects, we assume the easy axis anisotropy direction is distributed among a length scale defined by a "grain" size. The grains vary in size taking an average diameter of $D_G = 10$ nm. The direction of the uniaxial anisotropy ($\vec{u}_k$) of each grain is mainly directed along the perpendicular direction (z-axis) but with a small in-plane component which is randomly generated over the grains. The maximum percentage of the in-plane component of the uniaxial anisotropy unit vector ($\varepsilon$) is varied from 2% to 10% ($0.02 \leq \varepsilon \leq 0.10$). We have considered several different grain patterns generated randomly and it was confirmed that the presented results do not differ significantly from pattern to pattern.

Two strips with two different widths are studied here: $w = 384$ nm and $w = 3000$ nm. The strips are discretized using a finite difference scheme with cells composed of 3 nm × 3 nm × 1 nm: the thickness of the cell is the same with that of the CoFeB strip ($t_{FM} = 1$ nm). A micromagnetic study using the real dimensions of the experimental samples (~40 μm long wires and width of $w$~5000 nm) is not possible due to computer memory limitations. Therefore, we simulate strip lengths ($\ell$) of 3.072 μm and 6.144 μm.

**(b) Annihilation field of several DWs as a function of the DMI**

In order to support the experimental observations presented in Fig. 3 of the main text, micromagnetic simulations are performed for a strip with $\ell = 3.072\ \mu m$, width $w = 384$ nm and thickness $t_{FM} = 1$ nm with the same material parameters as stated above. Thermal fluctuations at room temperature and disorder defined by grains are taken into account to describe realistic conditions. The strip contains initially 31 domain walls (DWs), and a series of consecutive negative out-of-plane fields $H_z$ are applied, each one for a duration ($t_H$) of 10 ns. The number of DWs as a function of the applied field $H_z$ is plotted in Fig. S1(a). Snapshots of the magnetic state of the strip for representative fields are depicted in Fig. S1(b). The number of DWs remains constant in the range $0 < |H_z| <$ 35 Oe, but the domains magnetized in the opposite direction to the field (white color) shrink and the ones magnetized parallel to the field (black color) widen. As $H_z$ increases in magnitude, pairs of DWs start to collapse, and the number of DWs decreases. The field needed to annihilate all the walls is $H_{AN} \sim 90$ Oe. It was verified that this annihilation field ($H_{AN}$) and the dependence of the number of DWs on the applied field $H_z$ do not significantly vary for other randomly generated grain patterns with the same characteristics (average grain diameter $D_G = 10$ nm, $\varepsilon = 0.02$). Similar number of DWs vs $H_z$ curves are also obtained when the field duration $t_H$ and/or the strip width $w$ are increased to $t_H = 100$ ns and $w = 3000$ nm. These micromagnetic results qualitatively support the experimental observations presented in Fig. 3 of the main text.

The same numerical study is also performed with different DM exchange constant ($D$). The micromagnetic results are shown in Fig. S1(c), which indicate that the annihilation field $H_{AN}$ monotonically increases with $D$.

**(c) Distance between two DWs as function of the out-of-plane field**

In order to characterize the repulsive force between chiral walls we performed micromagnetic studies on the distance ($d_{DWs}$) between two DWs as function of the out-of-plane field $H_z$. We consider a strip with $\ell = 6.11$ µm, $w = 384$ nm and $t_{FM} = 1\ nm$ and the same material parameters as above. A down-up-down (↓↑↓) magnetic configuration with two domain walls initially separated by $d_{DWs}(t = 0) \sim 1$ µm is considered. A series of negative out-of-planed fields $H_z$ with increasing magnitude is applied ($H_z: 0, \ldots, -100$ Oe ). Each field is applied for a duration of $t_H = 50$ ns before changing the value to the next in the series. The terminal distance between the two walls ($d_{DWs}(t = t_H)$) is computed for each $H_z$. Micromagnetic results are shown in Fig. S2(a) under several circumstances. For the case of an ideal strip without defects (black solid squares), the distance between walls monotonically decreases as the magnitude of the field $|H_z|$ increases, and the DWs collapse when the field reaches the annihilation threshold $|H_{AN}| \sim 80$ Oe. This ideal case is perfectly reproduced by the analytical dipolar field $H_{d,DD}$, (see Ref.[11]), which is presented by the solid green line in Fig. S2(a). This dipolar field $H_{d,DD}$ accounts for the magnetostatic interaction of the side domains (magnetized down, ↓) with the central one (magnetized up, ↑). $H_{d,DD}$ is given by[11]

$$H_{d,DD} = -\frac{M_s}{\pi}\left(2\operatorname{atan}\left(\frac{2wd_{DWs}}{t\sqrt{t^2+w^2+4d_{DWs}^2}}\right) - \operatorname{atan}\left(\frac{(\ell-d_{DWs})w}{t\sqrt{t^2+w^2+(\ell-d_{DWs})^2}}\right) - \operatorname{atan}\left(\frac{(\ell+d_{DWs})w}{t\sqrt{t^2+w^2+(\ell+d_{DWs})^2}}\right)\right) \quad (S3)$$

The distance between walls as function of the applied field is also evaluated under realistic conditions considering grains. At zero temperature (open blue circles in Fig. S2(a)), the distance ($d_{DWs}$) remains constant in a field range of $0 < |H_z| < 10$ Oe. This is due to the disorder imposed by the grains which introduces a pinning threshold field (or a propagation field $|H_P|$, i.e. $|H_P|$ at 0 K) acting against the free DW propagation. Note that the same propagation field is also obtained from the analysis of the field-driven dynamics of a single DW in a strip with the same grain pattern (not shown). For larger fields ($|H_z| > |H_P| = 10$ Oe), the distance decreases monotonically, and for large enough fields ($|H_z| > 15$ Oe), a similar dependence of $d_{DWs}$ vs $H_z$ is found with that of the defect-free case. Note that the annihilation field does not differ significantly from that obtained for the defect-free strip ($|H_{AN}|\sim 80$ Oe).

At room temperature, there is no null probability of DW propagation under fields smaller than the propagation field ($|H_z| < |H_P|$), and the distance between walls decreases monotonically as $|H_z|$ increases (red solid circles in Fig. S2(a)). Even under these realistic conditions the dependence of $d_{DWs}$ vs $H_z$ is similar to that obtained from the defect-free strip at zero temperature. The annihilation field ($|H_{AN}|\sim 75$ Oe) is slightly reduced due to thermal activation with respect to the zero temperature case.

The repulsion between chiral DWs can be estimated by applying an out-of-plane field that compresses the center domain. As it can be easily imagined, the repulsion between chiral DWs depends on their relative distance. For the system numerically evaluated here, the repulsion is negligible for relative distance of $d_{DWs} \gtrsim 1$ μm as it can be inferred from Fig. S2(a). However, the repulsion becomes relevant for smaller distances.

Figure S2(b) shows the separation distance between the two walls against the DM exchange constant ($D$). Thermal effects and realistic disorder are taken into account with a fixed grain pattern. As observed in Fig. S2(b) for $D < 0.24$ erg cm$^{-2}$ or $d_{DWs} \gtrsim$ 50 nm, analytical prediction for the dipolar field between domains[11] $H_{d,DD}$ (Eq. (S3)) agrees well with the micromagnetic results, i.e. the separation distance due to $H_{d,DD}$ is in accordance with that from the micromagnetic simulations. $H_{d,DD}$ is represented by the green line in Fig. S2(b).

However, for larger values of $D$ smaller separation distance ($d_{DWs}$) between walls can be achieved under stronger out-of-plane fields. The micromagnetic results deviate from the analytical prediction given by Eq. (S3) for larger applied field ($|H_z| > {\sim}80$ Oe) (see Fig. S2(b) for $D = 0.48$ erg cm$^{-2}$, blue triangles). In this high field range the distance between walls approaches $d_{DWs} \approx 50$ nm, and the behavior of $d_{DWs}$ vs $H_z$ is better described by the dipolar field between the magnetic moments of the walls ($H_{d,MM}$)[12], which can be can be expressed as

$$H_{d,MM} = \frac{3 M_s \pi^2 \Delta^2 t w}{8 \pi d_{DWs}^4} (2 \sin\psi_L \sin\psi_R - \cos\psi_L \cos\psi_R) \qquad (S4)$$

where $\Delta$ is the DW width and $\psi_L$ and $\psi_R$ are the angles of the internal magnetization with respect to the *x*-axis within the left and the right walls, respectively. $H_{d,MM}$ is represented by the blue line in Fig. S2(b). Note that this dipolar interaction $H_{d,MM}$ scales with $1/(d_{DWs})^4$, and therefore, it becomes significant only when the walls are very close to the each other. This short length repulsion increases with $D$. As a consequence, the annihilation field $H_{AN}$ increases with $D$ (see Fig. S2(c)).

Snapshots of the magnetic state for a field $|H_z|$ just below the annihilation threshold ($|H_{AN}(D)|$) are shown in Fig. S2(d) for different values of the DM exchange constant ($D$). For achiral walls ($D < D_c {\sim} 0.2$ erg cm$^{-2}$ with the parameters of the present analysis) the walls adopt an antiparallel Bloch configuration (see Fig. S2(d) for $D = 0$ and $D = 0.12$ erg cm$^{-2}$ cases, where $\psi_L \sim 90^0$ and $\psi_R \sim 270^0$), and the annihilation field $H_{AN}$ is below the crossing of the analytical curves $H_{d,DD}$ and $H_{d,MM}$. For higher values of $D$ ($D \geq D_c \sim 0.2$ erg cm$^{-2}$) the internal magnetic moments depict a chiral Néel configuration with $\psi_L \sim 0^0$ and $\psi_R \sim 180^0$ (see Fig. S2(d) for $D = 0.24$ erg cm$^{-2}$ and $D = 0.48$ erg cm$^{-2}$ cases). These internal moments generate an additional repulsive force[12] between walls which becomes dominant at very short distances ($d \lesssim 50$ nm). As the chiral Néel configuration with $\psi_L \sim 0^0$ and $\psi_R \sim 180^0$ becomes more stable as the $D$ increases, the applied field needed to collapse the walls increases with $D$, explaining the dependence of the annihilation field with $D$.

### (d) DW velocity of uncoupled and coupled DWs

With the aim of explaining the experimental results of Fig. 4 in the main text, the current-driven domain dynamics are evaluated under the presence of static out-of-plane field $H_z$. We first focus on the case of a single DW in a strip with $\ell = 6.11$ µm, $w = 384$ nm and $t_{FM} = 1$ nm. Realistic conditions (grains and thermal fluctuations at room temperature) are considered for these simulations. Current pulses ($\vec{J}(t)$) with zero rise and fall times are instantaneously applied at $t = 0$ with a fixed amplitude of $J = 0.2 \times 10^8$ A/cm$^2$ and a pulse length of $t_p = 10$ ns. The total temporal window is $t_w = 50$ ns, and the field $H_z$ is statically applied during the whole temporal window. The temporal evolution of the magnetization is numerically computed by solving Eq. (S1), and the average DW velocity as a function of $H_z$ is obtained for each wall from the total distance $[q(t_w)]$ the wall travelled during the time window $t_w$ divided by the pulse length, i.e.

$v = q(t_w)/t_p$. Micromagnetic results for both isolated up-down (↑↓) and down-up (↓↑) DWs are presented in Fig. S3(c). Similar to the experimental results of Fig. 4(a) in the main text for a single DW, the DW velocity scales linearly with $H_z$. These experimental results (single domain wall case) are also replicated in Fig. S3(a) to facilitate the comparison with the µM results of Fig. S3(c). Similar results are also obtained from µM simulations using a strip with two DWs placed very far apart, that is, when their initial relative distance is $d_{DWs} \sim 1$ µm (results not shown).

With the aim of describing the DW motion of coupled walls, we set the initial distance between the two walls by applying a static field $H_z$: the separation distance $d_{DWs}$ are shown in Fig. S2(a). From this initial state for each field $H_z$, both the current pulse $J(t)$ and the static field $H_z$ are applied at $t = 0$, and the dynamics of the down-up (↓↑, at left side) and the up-down (↑↓, at right side) walls are evaluated by numerically solving Eq. (S1). These micromagnetic results are shown in Fig. S3(d), which depict the same trend with that of the experimental observations of Fig. 4(a) in the main text for the case of coupled walls. (The experimental results (coupled domain walls case) are also replicated in Fig. S3(b)). Both ↓↑ and ↑↓ walls move with the same velocity for negative fields $H_z < -5$ Oe. Note that the terminal distance $d_{DWs}(t_p)$ between the two walls for this field range is ~100 nm, where the coupling between walls becomes relevant. As the field increases from $H_z = -4$ Oe to $H_z = 0$, the compressing force of the field ($H_z$) cannot balance the repulsion between walls and therefore the walls are no longer coupled: $d_{DWs}(t_p)$ increases with increasing $H_z$. As a consequence, the velocity of the two walls becomes different even at zero field. The difference in the velocities of ↓↑ and ↑↓ walls is due to the different force exerted by the field and current: the current pulse drives both walls with the same effective force but the negative field ($H_z < 0$) promotes different velocities as it pushes the left ↓↑ wall towards to the right, while the same field pushes the right ↑↓ to the left. When the field polarity reverses (positive fields $H_z > 0$), separation of the two walls is promoted by the field $H_z (> 0)$. A positive current pulse ($J > 0$) moves both walls to the right. However, $H_z > 0$ drives the right ↑↓ wall to the right but the left ↓↑ wall is pushed to the left. This explains the increase of the ↑↓ wall velocity for positive current pulses and positive fields, and also the corresponding decrease of the ↓↑ wall velocity.

**S2. 1D model for two domain walls**

**(a) Model description**

In order to further support the experimental observations and the micromagnetic simulations, DW dynamics under current pulses and static out-of-plane field are evaluated in the framework of a 1 dimensional model (1D) extended for two DWs and taking into account the relative dipolar interaction between them[11, 12]. Such domain wall dynamics are described using the following two time dependent variables: the wall position $q_i(t)$ and the wall magnetization angle $\psi_i(t)$. The four coupled differential equations, two for each wall, are[12]

$$\frac{dq_i}{dt} = \frac{\Delta}{(1+\alpha^2)}\left(Q_i \Omega_{A,i} + \alpha \Omega_{B,i}\right) \tag{S5a}$$

$$\frac{d\psi_i}{dt} = \frac{1}{(1+\alpha^2)}\left(-\alpha \Omega_{A,i} + Q_i \Omega_{B,i}\right) \tag{S5b}$$

where the index $i: L, R$ corresponds to the left and the right DWs within the strip, and

$$\Omega_{A,i}(\psi_i, \psi_j, d) = -\frac{1}{2}\gamma_0 H_K \sin(2\psi_i) - \frac{\pi}{2}\gamma_0 Q_i H_D \sin(\psi_i) \tag{S6}$$

$$\Omega_{B,i}(\psi_i, \psi_j, d) = \gamma_0 Q_i H_{z,i} + \frac{\pi}{2}\gamma_0 Q_i H_{SH} \sin(\psi_i) - \frac{\pi}{2}\gamma_0 H_{FL} \sin(\psi_i) \tag{S7}$$

with $Q_i = +1$ and $Q_i = -1$ represent ↑↓ and ↓↑ walls, respectively. The magneto-static anisotropy field associated with the wall is expressed as $H_K = \frac{4 t_{FM} M_S \log(2)}{\Delta}$ [4, 13], where $M_S$ is the saturation magnetization, $\Delta = \sqrt{A/K_{eff}}$ is the domain wall width parameter and $t_{FM}$ is the thickness of the magnetic layer. $\alpha$ and $K_{eff} = K_u - 2\pi M_S^2$ are the Gilbert damping parameter and the effective magnetic anisotropy constant of the magnetic layer, respectively. $H_{DM}$ and $H_{SH}$ are the Dzyalonshinskii-Moriya (DM) offset field and the spin Hall effective field, respectively. $H_{DM}$ and $H_{SH}$ can be explicitly written as $H_{DM} = \frac{D}{\Delta M_S}$ and $H_{SH} = -\frac{\hbar \theta_{SH}}{2 e M_S t_{FM}} J$, where $D$ is the DM exchange constant, $\theta_{SH}$ is the spin Hall angle of the heavy metal layer and $J$ is the current density that flows into the heavy metal layer. The field like contribution is $H_{FL} = k H_{SH}$. The definitions of the constants used here are: $\gamma$ is the gyromagnetic ratio, $\hbar$ is the reduced Planck constant and $e$ is the electron charge. The out of-plane field $H_{z,i}$ consists of two contributions: the externally applied magnetic field $H_z$ and the dipolar field $H_d = H_{d,DD} + H_{d,MM}$, where $H_{d,DD}$ is the magnetostatic interaction of the lateral domains on the central one[11], given by Eq. (S3), and $H_{d,MM}$ is the interaction between the internal magnetic moments inside of the walls[12], given by Eq. (S4).

**(b) Comparison to micromagnetic results**

We have performed an analysis of the DW dynamics under current pulses and static out-of-plane fields using the 1D model by numerically solving Eq. (S3)-(S7). The strip dimensions, material parameters and the current pulse amplitude and length are the same as those for the micromagnetic study. The corresponding 1D results of the DW velocity as a function of the field are shown in Fig. S3(e)-(f). As it can be observed, similar trends are obtained. Small quantitative discrepancies are observed between µM and 1D results; however this is expected since, contrary to the realistic micromagnetic modelling, the 1D results are obtained for a perfect strip at zero temperature. The dependence of the terminal distance between the walls ($d_{DWs}$) on the applied field $H_z$ is shown in Fig. S3(g), where again a good agreement between realistic µM and 1D is achieved.

**Supporting references**

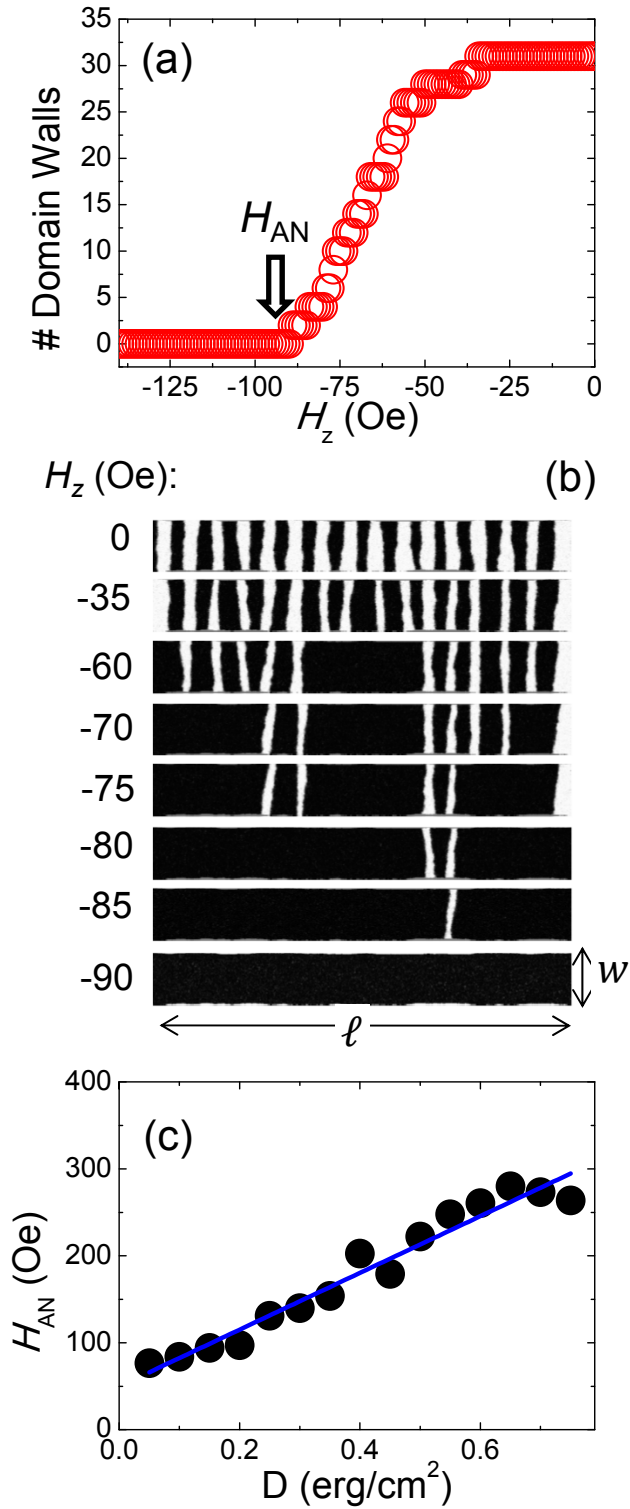

**Fig. S1. Estimation of repulsive interaction using micromagnetic simulations.** (a) Variation of the number of domain walls as a function of out of plane magnetic field ($H_z$) calculated using micromagnetic simulations with $D =$ 0.24 erg cm$^{-2}$. The width of the wire is $w = 384$ nm and its length is $\ell = 3.072$ μm. The length of the field pulse $t_H =$ 10ns. The calculation starts from 31 domain walls each separated by ~100 nm as the initial state. (b) Typical snapshots of the magnetization for several applied fields. (c) The calculated average $H_z$ needed to annihilate all domain walls plotted against the DM exchange constant ($D$). The blue line is linear fit to data.

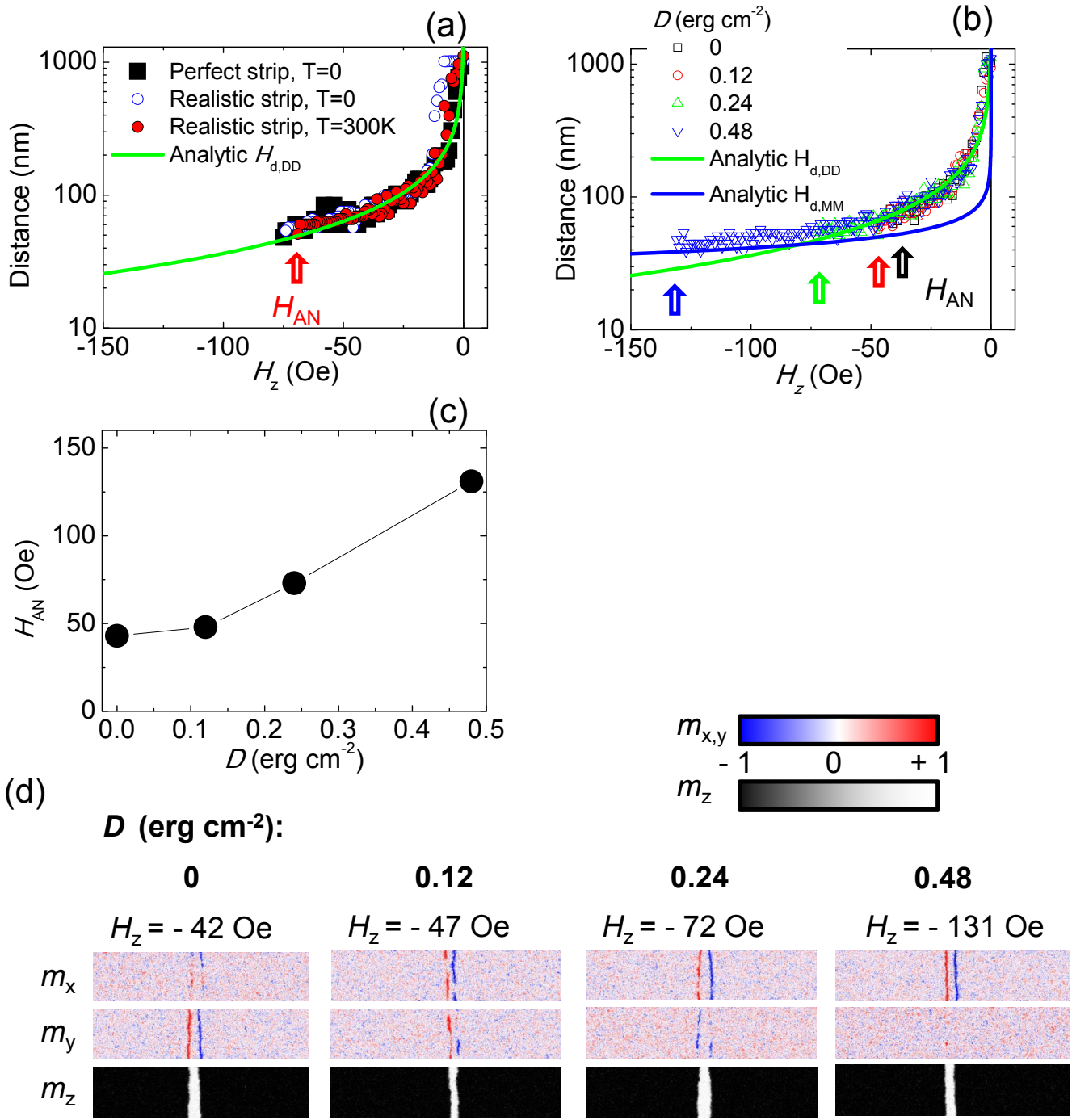

**Fig. S2. Calculated separation distance between the two domain walls.** (a) The separation distance of two domain walls plotted as a function of out of plane magnetic field ($H_Z$). The symbols indicate estimation of the distance using micromagnetic simulations. Black squares correspond to a defect-free case at zero temperature. Circles correspond to a realistic strip with grains (pinning): open and solid circles were obtained at zero and at room temperature respectively. The solid green line represents analytical calculation using Supplementary eq. (3). (b) Micromagnetic results for different values of the DMI parameter ($D$) considering realistic conditions with disorder and thermal fluctuations at room temperature. The solid green and blue lines represent analytical calculation using Supplementary eq. (3) and (4) respectively. (c) Annihilation field ($H_{AN}$) of the two walls for different values of the DMI parameter ($D$). (d) Micromagnetic snapshots computed at the field before annihilation threshold for different values of the DMI parameter ($D$).

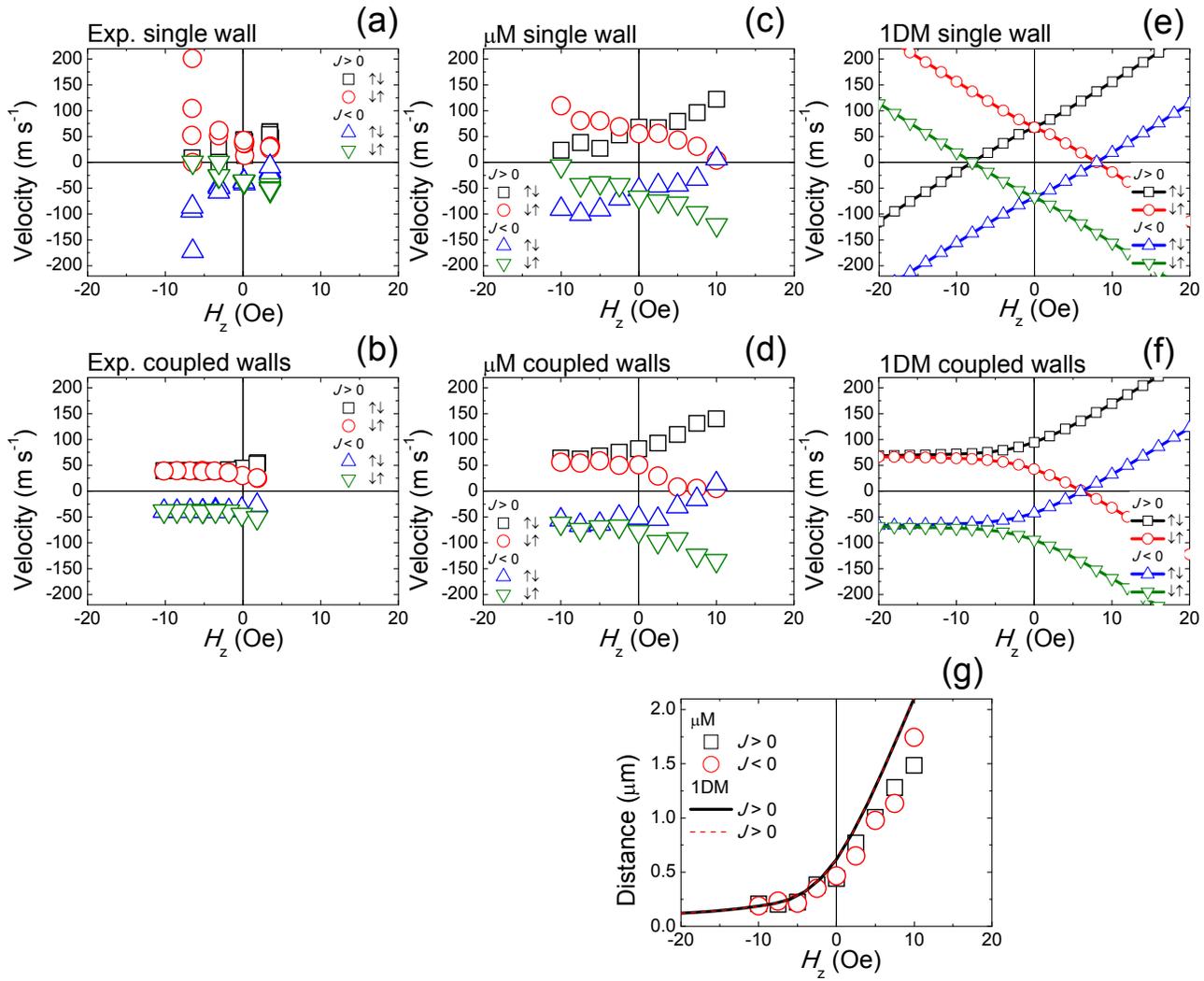

**Fig. S3. Current driven motion of uncoupled and coupled chiral walls.** Experimental (Exp.) results of the domain wall velocity as a function of the applied field $H_z$ for uncoupled (a) and coupled walls (b). These results are the same as in Fig. 4 of the main text. (c) and (d) are micromagnetic (µM) results of the domain wall velocity as a function of the applied field $H_z$ for uncoupled and coupled walls respectively. (e) and (f) are results obtained using the 1D Model (1D) for the domain wall velocity as a function of the applied field $H_z$ for uncoupled and coupled walls respectively. See further details in the text. A current pulse with $J = 0.2 \times 10^8$ A/cm$^2$ and $t_p = 10$ ns was applied for each field $H_z$, which is statically applied during the full temporal window $t_w = 50$ ns. These results were obtained considering a realistic strip (pinning) at room temperature. The terminal distance ($d_{DWs}$) between the two initially coupled walls is plotted in (g), where both micromagnetic (µM) and 1D model results (1D) are shown together for both positive and negative current pulses.